\documentclass[showpacs,aps,amsmath,amssymb,twocolumn]{revtex4}

\usepackage{graphicx}
\usepackage{dcolumn}
\usepackage{bm}

\begin{document}

\title{Geometric Phases in Graphitic Cones}

\author{Claudio Furtado and Fernando Moraes}
\affiliation{Departamento de F\'{\i}sica, CCEN,  Universidade Federal 
da Para\'{\i}ba, Cidade Universit\'{a}ria, 58051-970 Jo\~ao Pessoa, PB,
Brazil}

\author{~A. ~M. de ~M. Carvalho }
\affiliation{Departamento de F\'{\i}sica, Universidade Estadual de Feira de Santana,
BR116-Norte, Km 3,
44031-460, Feira de Santana, BA, Brazil}

\begin{abstract}
In this article we use a geometric approach to study geometric phases in graphitic cones. The spinor that describes the low energy states near the Fermi energy acquires a phase when transported around the apex of the cone, as found by a holonomy transformation. This topological result can be viewed as an analogue of the Aharonov-Bohm effect. The topological analysis is extended to a system with $n$ cones, whose resulting configuration is described by an effective defect. 

\end{abstract}


\maketitle
\section{Introduction}
The electronic and structural
properties of materials made of curved monolayers of graphite are object of intensive research. Typical examples of these structures are found in 
fullerenes \cite{kroto},
nanotubes \cite{iij,min}, graphitic cones \cite{kris}, scrolls \cite{viculis}, etc...
Graphite is a network of Carbon 6-atom rings with a planar geometry, a regular tiling of the plane with hexagons. Referring to graphite as the defect-free system, two-dimensional structures of Carbon  with rings with varying number of atoms are full of defects, each one corresponding to a polygon of $n$ vertices, with $n\neq 6$.   The disclinations, topological defects associated to these Carbon rings  are responsible for the local curvature that bends those structures into their various shapes.

The defects can be conceptually generated by a ``cut and glue'' 
process, known in the literature as the Volterra process \cite{kleman}. 
In the case of graphite, if a sector of $60^{o}$,  measured from the center of one of the hexagons to two near-neighbors Carbon atoms, is removed, and the remaining structure 
is continuously bent until the  surface reconnects,  a graphitic nanocone is obtained. It is essentially a single pentagon surrounded by a host of hexagons. Conversely, a whole $60^{o}$ sector may be added to a hexagon, transforming it into a heptagon surrounded by hexagons only. The graphitic cones
are classified by their opening angle, or equivalently,
the number, $n_{\Omega}=6-n$, of removed sectors.

In this work we use the continuous approach to  topological defects in solids of Katanaev and Volovich \cite{kat}, which have shown the equivalence between
three-dimensional gravity with torsion and the theory of defects in solids. In this
approach the defects are represented by metrics that are solutions
of Einstein-Cartan equations in three space dimensions.  An application of this model to graphite has appeared in \cite{local}, where localization of charge near heptagonal rings has been verified.
In this geometrical background we analyze the global properties of the massless Dirac equation that describes the electronic spectrum  of graphite at the Fermi level  \cite{prb:div}. 

It was shown recently by Lammert and Crespi \cite{crespi} that a continuous theory for graphitic 
cones describes the topological aspects of their physical properties.
They proposed a continuous model for this
material using the  ``tight-binding'' approach. They used two sub-lattices to
describe graphene and got the following Hamiltonian for the system
\begin{equation}\label{hamilt}
H_{0}=-iv_{f}(\sigma_{1}\partial_{x} + \sigma_{2} \partial_{y}),
\end{equation}
where the $\sigma$'s are Pauli's spin matrices. They applied the
appropriate boundary conditions in the above Hamiltonian and
showed that the graphitic monolayer exhibited an Aharonov-Bohm\cite{pr:aha}
analogous effect.  Recently,  a series of papers have investigated curved carbon structures using
a massless Dirac equation to analyze the electronic properties of this structure.
Gonz\'alez, Guinea and Vozmediano \cite{voz1,voz2}
investigated fullerene molecules using the massless Dirac equation in the presence
of a magnetic monopole of half-integer charge in the geometry of the sphere. The electronic structure
of a graphitic nanoparticle was  investigated by Osipov, Kochetov and Pudlak \cite{osi1,osi2,osi3} using a field theory model. In this work, Osipov and co-workers investigated  disclinations in the  cone, sphere and hyperbolic geometries. The local density of states was investigated using the Dirac equation in these geometries.
\section{Geometric Approach to a Graphitic Nanocone}
In this section, we make use of the geometric theory of defects \cite{kat,fur} to describe disclinations
in graphenes and nanocones. In this theory, the modifications
introduced in the elastic continuum by the defect
are described by a metric. This metric that describes the medium
surrounding the defect is a solution of the three-dimensional
Einstein-Cartan equation and contains all topological information 
about the medium. Using this theory we will show that the geometric Aharonov-Bohm
effect appears naturally in this system, because of its
topological and geometrical properties.
The conical graphene leafs will be described in this
continuum approximation by the two-dimensional metric
\begin{eqnarray}
\label{dis}
ds^{2} = dt^{2}-d\rho^{2} - \alpha^{2}\rho^{2}d\phi^{2}, \label{cone_t}
\end{eqnarray}
where $\alpha$ is the deficit or excess angle. $\alpha$ is related to 
the angular sector $\lambda$, which is the sector that is removed
or inserted to form the defect, by the expression
$\alpha=1\pm \lambda/2\pi$. We also can relate the deficit/excess
angle to the number of sectors removed from the graphitic  monolayer 
in the following way
\begin{equation}\label{N}
\alpha= 1-\frac{n_{\Omega}}{6} .
\end{equation}

Values of $\alpha$ in the interval, $0<\alpha<1$, mean that 
we remove a sector of the leaf to form a defect, and in the
interval, $0<\alpha<1$, mean that we insert a sector in the leaf.
In this way, the opening angle of the wedge  is $n\frac{\pi}{3}$, with $0<n<6$. this disclination angle is related to the opening angle of the cone by $\theta=2 \sin^{-1}(1-n/6)$.  The graphene sheet is associated to $n=6$. The carbon nanocones are characterized by $0<n<6$. Cone angles $\theta= 19^{o}, 39^{o}, 60^{o}, 85^{o}$ and $113^{o}$ have been observed experimentally \cite{nat:kri}. Note that these structures represent cones of positive curvature. Graphitic cones characterized by a carbon heptagon have negative curvature and are obtained by a insertion of a angular sector in the carbon sheet.


\section{Holonomy in the Conic Geometry}
Now we consider a two dimensional conic geometry that describes a disclination in graphitic material. We employ parallel transport to investigate the geometric phase\cite{berry} of the electronic  wave function, in the context of the Lammert-Crespi model, in the geometry given by (\ref{cone_t}).
In order to study global properties of the geometry of a nanocone of graphite we compute the orthonormal frame matrix for the parallel transport of a spinor along a closed path, obtaining the  holonomy matrix. When a vector is parallelly propagated along a loop in a manifold $\cal{M}$, the curvature on the manifold causes the vector initially at $p\in \cal{M}$, to appear rotated with respect to its initial orientation in tangent space $T_{p}\cal{M}$, when it returns to $p$.The holonomy is the path dependent linear transformation responsible for this rotation.  Positive and negative curvature manifolds, respectively, yield deficit or excess angles between initial and final vector orientation, under parallel transport around such loops.  The holonomy is defined by the following expression \cite{valdir}
\begin{equation}\label{holo}
U(C)={\cal P}\exp \left(-\oint \Gamma_{\mu}(x) dx^{\mu} \right),
\end{equation}
where $\Gamma_{\mu}(x)$ is the spinorial connection. 

In the Lammert-Crespi model the electronic properties of the graphene are obtained from a Dirac Hamiltonian for massless fermions in $(2+1)$-dimensions. The wave function of the system is therefore described by a spinor. Hence, we investigate the parallel transport of these spinors in a conical space in order to obtain global properties of the model. We now introduce the appropriate dual $1-$form basis (co-frame) which describes the background of a graphitic nanocone, defined by $e^a=e^{a}_{\mu}dx^{\mu}$, where
\begin{subequations}
\begin{eqnarray}
e^{0}&=& dt\\
e^{1}&=& \cos(\phi) d\rho -\alpha \rho \sin(\phi)d\phi \\
e^{2}&=& \sin(\phi) d\rho +\alpha \rho \cos(\phi)d\phi .
\end{eqnarray}
\end{subequations}
The metric and the inverse tetradic fields are related by
\begin{equation}
E^{\mu}_{a}E^{\nu}_{b}\eta^{ab}=g^{\mu\nu},
\end{equation}
where $e^{a}_{\mu}E^{\mu}_{b}=\delta^{a}_{b}$.
The connection forms are obtained from the first of the Maurer-Cartan structure equations: 
$de^{a}+\omega ^{a}_{b} \wedge e^{b}=0$.
So, for the above $1$-forms we get the following connections 
\begin{equation}
\omega^{2}_{1}=-\omega^{1}_{2}=(\alpha -1)d\phi.
\end{equation}
The spin connections and the 1-form connections are related by: $\Gamma_{\mu}=\omega^{a}_{b}dx^{\mu}$. Henceforth, we obtain the following matrix connection
\begin{eqnarray}
\Gamma_{\mu}\left(
\begin{array}{ccc}
0 & 0 & 0\\
0 & 0 & -(\alpha-1) \\
0 & (\alpha-1) & 0
\end{array}
\right).
\end{eqnarray}
This enables us to determine the spinorial connections, which are described by
\begin{equation}
\Gamma_{\mu}(x)=-\frac{1}{4}\Gamma^{\alpha}_{\beta\,\mu}\gamma_{\alpha}\gamma^{\beta},
\end{equation}
where $\gamma^{\alpha}$ are Dirac matrices. Therefore, the spinorial connections for the cone are given by
\begin{equation}
\label{spinor}
\Gamma_{\mu}(x)= -i \frac{(\alpha -1)}{2}\sigma^{3}.
\end{equation}

The holonomy matrix, $U(C)$, that stands for parallel transport
of a spinor along a path $C$ around the cone, is given by
\begin{equation}
U(C)=\exp\left[\frac{-i}{2}(\alpha -1)\sigma^{3}\phi\right]_{0}^{2\pi},
\end{equation}
which can be expanded as
matrix as
\begin{equation}
U(C)=\cos[( \alpha-1)\pi] +i \sigma^{3}\sin[(\alpha -1)\pi].
\end{equation}
In terms of  $n_{\Omega}$, we have
\begin{equation}
U(C)=\cos[\frac{ n_{\Omega}\pi}{6}] -i \sigma^{3}\sin[\frac{
n_{\Omega}\pi}{6}].
\end{equation}
This expression gives the quantum phase acquired by the
wave function when transported around the nanocone.
That is, when we transport the wave function, $\Psi$, around the 
defect we obtain 
\begin{equation}
\Psi'=U(C)\Psi,
\end{equation}
where $U(C)$ provides the phase obtained by the wavefunction in the process. This result can be interpreted as a analogue Aharonov-Bohm effect \cite{alex,tod,burgers}, where the magnetic flux is replaced by a ''curvature flux". This result can, in principle, be verified
experimentally in these graphitic materials. 

The holonomy transformation is equivalent to a geometric phase. This statement can be directly verified by writing the massless Dirac equation in curved space 
\begin{equation}
\label{dirac}
i\gamma^{a}E^{\mu}_{a}D_{\mu} 
\psi=0,
\end{equation}
where $D_{\mu}$ is the covariant derivative for a spinor
\begin{equation}
D_{\mu}=\partial_{\mu}+ \Gamma_{\mu}(x).
\end{equation}
For metric (\ref{cone_t}), Dirac equation can be written as 
\begin{equation}
\label{dirac-x}
\left\{
i
\left[
\gamma^{0}\partial_{t}+\gamma^{\rho} \left( \partial_{\rho}-
\frac{1-\alpha}{\alpha \rho}
\right)
+\gamma^{\phi}\left(\frac{1}{\alpha \rho}\partial_{\phi}
\right)
\right]
-m
\right\}
\psi=0
\end{equation}
The solution of this equation  can be writen as 
\begin{equation}
\psi=\exp\left[\frac{-i}{2}(\alpha -1)\sigma^{3}\phi\right]\psi_{0},
\end{equation}
which leads to
\begin{equation}
\label{dirac}
i\gamma^{\mu}\partial_{\mu} 
\psi_{0}=0.
\end{equation}
This result allow us to interpret, $\exp \left(\int \Gamma_{\mu}(x) dx^{\mu} \right)$, as a geometric phase. The spinor acquires a nontrivial phase when moved
around the apex of the cone. This is a consequence of the the fact that the intrinsic curvature of the defect is accumulated in the apex of the cone. In the lattice description, the spinor acquires a nontrivial phase due the accumulated curvature in the pentagon sites of the nanocone structure. In the continuum limit approximation that we use the holonomy matrix gives the information about the geometric phase due to the parallel transport of a spinor in the carbon nanocone structure.    


\section{multicones}
In recent articles, the energetic stability of multiconic  structures of graphite  have been investigated \cite{prl:cha,prb:ber}. In this section, we use the geometric tools developed for a simple cone to investigate these multiconic Carbon structures. The holonomy transformation is therefore used to characterize the analogue Aharonov-Bohm effect in these systems. 

Let us consider a graphitic structure formed by many cones and
determine the correspondent holonomy transformations.
Initially, we consider a structure with only two cones
placed, respectively, at the points $\rho_{1}$ and $\rho_{2}$.
We then perform the parallel transport of along a closed path $C_{1}$ around the point $\rho_{1}$. Consequently, the holonomy (or the Dirac's factor) is given by
\begin{equation}
\psi^{(1)}(\rho,\phi)=\exp\left[-\frac{i}{2}(\alpha_{1}-1)\sigma^{3}
\int_{C_{1}}d\phi\right]\psi_{0}(\rho,\phi).
\end{equation}
After that, the parallel transport is made around the second cone,
following the contour $C_{2}$, which results in
\begin{eqnarray}
\psi^{(2)}(\rho,\phi)&=&\exp\left[-\frac{i}{2}(\alpha_{2}-1)\sigma^{3}
\int_{C_{2}}d\phi\right]\psi^{1}(\rho,\phi) \nonumber \\
&=&\exp\left[-\frac{i}{2}(\alpha_{1}+\alpha_{2}-2)\sigma^{3}
\int_{0}^{2\pi}d\phi\right]\psi_{0}(\rho,\phi).
\end{eqnarray}

Following the above reasoning, we generalize the result to $m$ cones placed, respectively,
at the points $\quad \quad \rho_{1},\rho_{2},\cdot\cdot\cdot,\rho_{m}$, to
\begin{equation}
\psi^{(1,2,\cdot \cdot \cdot, m)}(\rho,\phi)=\exp
\left[-\frac{i}{2}\sum_{j=1}^{m} (\alpha_{j}-j)\sigma^{3}
\int_{0}^{2\pi}d\phi\right]\psi_{0}(\rho,\phi).
\end{equation}
Since each disclination parameter $\alpha_{j}$ is related to the numbers
of removed sectors $n_{\Omega}^{j}$ by the expression
\begin{equation}
\alpha_{j}=1-\frac{n_{\Omega}^{j}}{6},
\end{equation}
we can write the holonomy transformations for $m$ cones as
\begin{equation}
U_{m}(C)=\exp \left( \frac{i}{2}
\sum_{j=1}^{m}\frac{n_\Omega}{6}\sigma^{3}\int_{0}^{\pi}d\phi'\;
\right),
\end{equation}
which results in 
\begin{equation}
\label{ncones}
U_{m}(C)=\cos\left( \sum_{j=1}^{m}\frac{n_{\Omega}\pi}{6}\right)-
i\sigma^{3}\sin \left( \sum_{j=1}^{m}\frac{n_{\Omega}\pi}{6}\right).
\end{equation}
We note in the expression (\ref{ncones}), that the problem
of $m$ cones can be replaced by one with an effective defect, with $n_{\Omega}$
given by
\begin{equation}
n_{\Omega}^{\mbox{eff}}=\sum_{j=1}^{m}n_{\Omega}^{i}.
\end{equation}
In this way, we can analyze the geometric phase for the multicone structure containing 
many isolated disclinations of various strengths.  The geometric phase for the spinor transported  around a multiconic sheet for assorted effective disclination angles are exhibited in Table I. The maximum numbers of removed sectors is $n_{\Omega}^{eff}=5$,
which corresponds to the topological condition for the existence
of nanocones of graphite. Note that the multicone approach  can be used to the limit situation of the  cylindrical nanotube formed by the removal of six  $\frac{\pi}{3}$ sectors. In this case the spinor is rotated by an angle of the $\pi$.  This approach also illustrates the power of the  holonomy transformation to calculate the geometric phase to multiconic structures. The direct calculation, via Dirac equation in the general metric that describes the medium, is much more difficult to solve. This method can be used to investigate  negative curvature graphene sheets formed when  angular sectors are inserted.

\section{Concluding Remarks}

In this work we investigate quantum phases in nanocones of graphite. We show 
that the phase of the wave function is of the Aharonov-Bohm type. We use the 
calculus of holonomy to describe the analogue Aharonov-Bohm effect in nanocones of graphite.
Comparing with the Aharonov-Bohm phase $\oint A_{\mu}dx^{\mu}$, the analogue AB 
effect occurs when the holonomy $\oint \gamma_{\mu}(x)dx^{\mu}$ is nontrivial. 
We also analyze the Aharonov-Bohm effect in a multiconic structure formed by a set
of disclinations of same or different strength. Finally, we suggest that geometric phase obtained in this work can be used in the future investigations of 
topological quantum computation in nanocone structures along the lines pointed  in the references\cite{pachos,pachos2}.

\begin{table}[h]
\begin{center}
\begin{tabular}{|c|c|c|c|c|c|} \hline 
Structure & $n_{\Omega}$ & $n_{\Omega}^{\mbox{eff}}$ & $U_{m}(C)$ &
$\phi^{\mbox{total}}$&$ \alpha^{eff}$\\\hline
$\lambda_{1}=\frac{\pi}{3}$ & $1$ & $1$ &
$\exp(-\frac{\pi\sigma^{3}}{3})$ & $2\pi -\frac{\pi}{6}$ & $\frac{5}{6}$ \\ \hline
$\lambda_{1}=\lambda_{2}=\frac{\pi}{3}$ & $1$ & $2$ &
$\exp(-\frac{\pi\sigma^{3}}{3})$ & $2\pi -\frac{2\pi}{3}$ &$\frac{2}{3}$\\ \hline
$\lambda_{1}=\lambda_{2}=\lambda_{3}=\frac{\pi}{3}$ & $1$ & $3$ &
$\exp(-\frac{\pi\sigma^{3}}{2})$ & $2\pi -\pi$ $\frac{5}{6}$&$\frac{1}{2}$ \\ \hline
$\lambda_{1}=\lambda_{2}=\lambda_{4}=\frac{\pi}{3}$ & $1$ & $4$ &
$\exp(-\frac{4\pi\sigma^{3}}{3})$ & $2\pi -\frac{4\pi}{3}$& $\frac{1}{3}$ \\ \hline
$\lambda_{1}=\lambda_{2}=\lambda_{4}=\lambda_{5}=\frac{\pi}{3}$ & $1$ &
$5$ & $\exp(-\frac{5\pi\sigma^{3}}{6})$ & $2\pi -\frac{5\pi}{3}$& $\frac{1}{6}$ \\
\hline
$\lambda_{1}=\frac{2\pi}{3},\quad \lambda_{2}=\frac{2\pi}{3}$ & $1$ &
$4$ & $\exp(-\frac{2\pi\sigma^{3}}{3})$ & $2\pi -\frac{4\pi}{3}$& $\frac{1}{3}$ \\
\hline
\end{tabular}
\end{center}
\caption{The conic structure of the effective disclination given by $2\pi \alpha^{eff} $. The geometric phase for the multicone formed by the removal of several sectors of strength $\lambda$ is given by $U(C)$.}
\end{table}


\end{document}